\documentclass[
aip, 
pop, 
amsmath,amssymb, 
reprint, 
floatfix
]{revtex4-1} 
\usepackage{graphicx} 
\graphicspath{{Figs/}}
\usepackage{bm} 

\usepackage[mathlines]{lineno}
\usepackage{comment}
\usepackage[colorlinks=true, allcolors=blue,breaklinks=true]{hyperref} 
\usepackage[all]{hypcap} 
\begin{document}
%
\title{Interactions Between Two Magnetized Plasma Pressure Filaments: Drift-Alfvén Modes and Transport}
%
%

\author{T. Simala-Grant}
\affiliation{Department of Physics, University of Alberta, Edmonton, Alberta T6G 2E1, Canada}

\author{S. Karbashewski}
\affiliation{Department of Physics, University of Alberta, Edmonton, Alberta T6G 2E1, Canada}

\author{R.D. Sydora}
\affiliation{Department of Physics, University of Alberta, Edmonton, Alberta T6G 2E1, Canada}

\author{B. Van Compernolle}
\affiliation{General Atomics, San Diego, California 92121, USA}

\date{\today} 
%
%
\begin{abstract}
A set of experiments on merging dynamics of localized magnetized plasma pressure filaments in close proximity has been carried out in a large linear magnetized plasma device. A two filament configuration is used to determine the separation scale of cross-field interaction. Fluctuation analysis of the ion saturation current confirms the presence of unstable drift-Alfv\'{e}n modes on each filament which are predominantly driven by the steep electron temperature gradient. When the filaments are separated a distance of approximately five times the size of a single filament or greater, no interaction is observed and the drift-Alfv\'{e}n modes remain uncoupled. When brought closer together, significantly stronger interaction occurs with transfer of charge and energy, leading to the generation of inter-filament electric fields. This has the effect of rotating the filaments and influencing the merging dynamics. When the filaments are asymmetrically heated, the cooler filament becomes wrapped around the other, and the drift-Alfvén modes become fully coupled in a global mode structure. Transport between the filaments is explained in terms of $\bf{E}\times \bf{B}$ drift and the local potential wells of the filaments, and drift-Alfvén mode coupling occurs when the filaments are brought within the distance of a collisionless skin depth.

\end{abstract}

\maketitle
%
%
%
\section{Introduction}

Mesoscale plasma structures known as blobs, which are filamentary structures extended along a background confining magnetic field, have become a central topic in transport physics since they have been observed in virtually all magnetic fusion devices~\cite{Boedo2003,Theiler2011,Terry2017}. The blobs were initially related to bursty or intermittent transport events in the edge region of toroidal plasmas and measured as large spikes of the ion saturation current collected by electrostatic probes. The filamentary nature of the blob was then confirmed by direct observations with fast cameras~\cite{Zweben2002,Zweben2016,Terry2017,Offeddu2022}. Blobs commonly refer to positive amplitude pressure perturbations in relation to the background pressure and their cross-field propagation constitutes a major transport mechanism for plasma density and energy~\cite{Krasheninnikov2008,DIppolito2011}.

In the case of laboratory magnetized plasmas, blobs can be generated nonlinearly from interchange/drift plasma turbulence~\cite{Garcia2004,Furno2008,Bisai2019}, thus forming multiple blobs in close proximity. 
They can further interact with each other leading to merging or breakup, which generates many structures with varying sizes. In addition, blobs are affected by internal instabilities that may induce fragmentation~\cite{Angus2012,Easy2014} and it has been shown that a competition between the time scales for instability and transport determines the range of allowed blob sizes and shapes, thus constraining the overall radial velocity.

Previous studies of blob-filament dynamics in linear magnetized plasma devices have been made where these structures and intermittent dynamics are associated with instabilities in drift waves, flow shear and rotational modes where centrifugal forces can mimic magnetic
curvature~\cite{Nielsen1996,Windisch2006,Antar2007}. Blob-filament structures have also been observed in rapidly rotating plasma
columns~\cite{Pierre2004} and in the shadow of a limiter placed inside the cylindrical column~\cite{Carter2006}.
The vast majority of experiments have focused on self-generated blob-filaments involving density filaments, while heat transport studies have been carried out using electron temperature filaments formed using a thermalized electron beam source in a linear plasma device~\cite{Burke1998,Sydora2019,Karbashewski2022}.

In this work we focus on the merging dynamics of two interacting blob-filaments where the internal pressure is dominated by the electron temperature. The symmetry breaking of the radial temperature gradient on each filament is impacted through filament-filament interaction, thus altering the convective and drift-Alfvén mode structure\cite{Sydora2024}. Another important aspect addressed in this work is the impact of asymmetric heating and merging of filaments with unequal electron temperature.

The organization of this paper is as follows. Section~\ref{sec:ExpSetup} describes the experiment setup and parameters used to initialize the filamentary structures. The results 
from Langmuir probe measurements are presented in Section~\ref{sec:ExpResults}, beginning with the filament evolution as a function of spatial separation. Cross-field maps of the electron temperature, density, and space potential at fixed axial position are shown. This is followed by a detailed characterization of the mode structure from the dominant frequencies and cross power/cross phase analysis. The degree of filament-filament coupling is quantified using a coherence measure.
Section~\ref{sec:Discussion} contains a discussion on the results and physical processes, while Section~\ref{sec:Summary} gives a summary of the main results.


%
\section{Experiment Setup}\label{sec:ExpSetup}

\begin{figure}
{\includegraphics[trim= 0cm 0cm 0cm 0cm,clip,width=\linewidth]{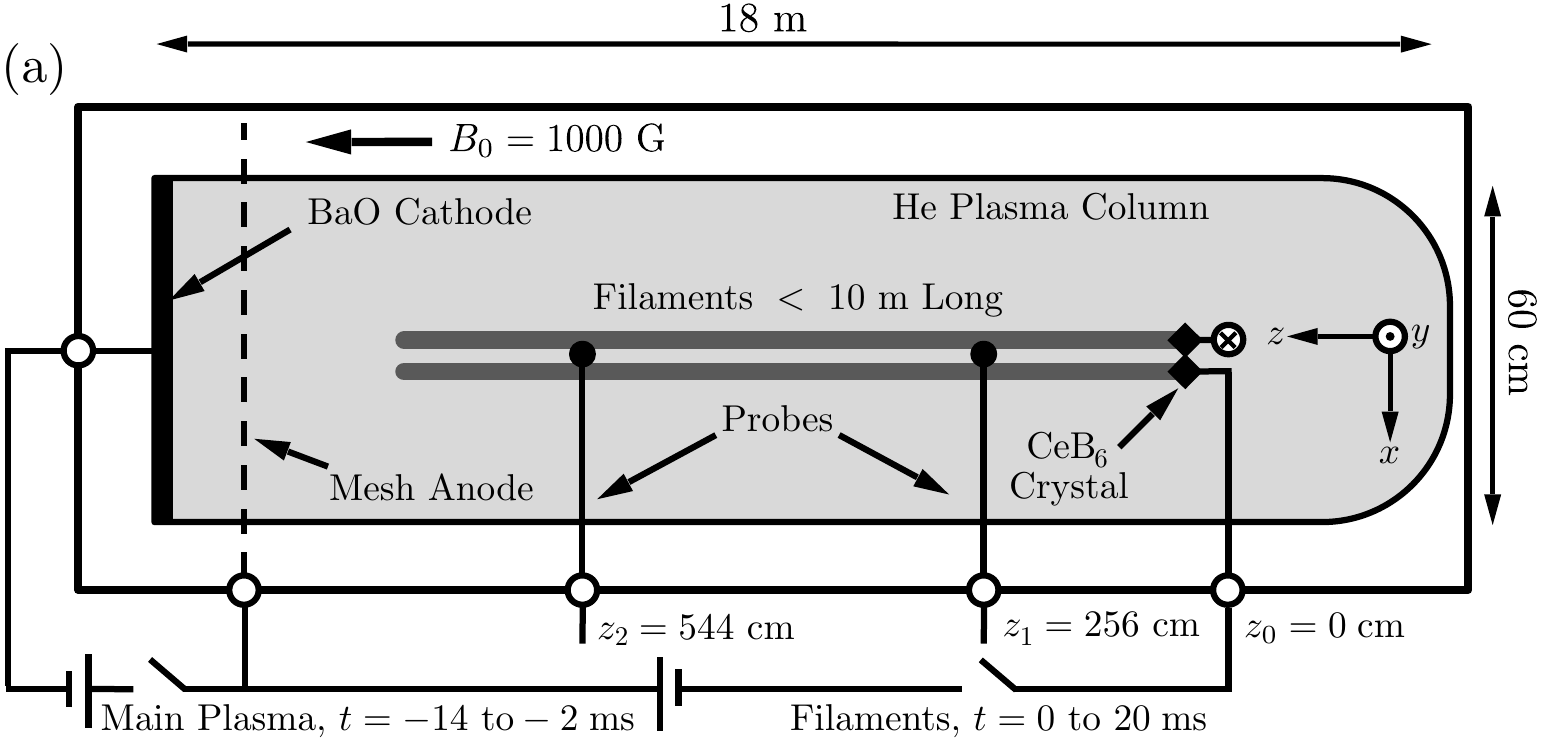}}
\caption{Schematic (not to scale) of the experimental setup on the LAPD. In the afterglow phase of the main discharge, the $\mathrm{CeB}_6$ cathodes inject electron beams into the plasma, creating temperature plumes that are transported down the length of the device and form narrow filaments of elevated temperature. Langmuir probe measurements are made at axial positions $z_1 = 256~\mathrm{cm}$ and $z_2 = 544~\mathrm{cm}$ away from the filament heat source.  \label{fig:Schematic}}
\end{figure}

We conducted the experiments on the upgraded Large Plasma Device (LAPD) at the Basic Plasma Science Facility (BaPSF) at the University of California, Los Angeles (UCLA)~\cite{Gekelman2016}. Fig.~\ref{fig:Schematic} presents a schematic of the LAPD and experimental setup. The LAPD produces a cylindrical, magnetized plasma column that is $18~\mathrm{m}$ long and $0.6~\mathrm{m}$ in diameter. Cylindrical electromagnets encircling the vacuum vessel provide the magnetizing field and are operated at approximately $0.1~\mathrm{T}$ ($1000~\mathrm{G}$) during our experiments. The plasma discharge is initiated by a cathode source located at one end of the device; at the time, a barium oxide (BaO) cathode was installed --- the LAPD has since been upgraded and now uses a lanthanum hexaboride (LaB$_6$) cathode. $0.5~\mathrm{m}$ from the cathode is a mesh anode biased at $70~\mathrm{V}$ above the cathode; this accelerates thermionic electrons into the chamber and ionizes the helium (He) gas filling the chamber at a fill pressure on the order of $10^{-4}~\mathrm{Torr}$. The main discharge lasts $12~\mathrm{ms}$ and produces a plasma with an electron temperature of $T_e\sim5~\mathrm{eV}$, density of $n_e\sim2\times10^{18}~\mathrm{m}^{-3}$, and ion temperature, $T_i$, less than $1~\mathrm{eV}$. When the main discharge ends, the plasma transitions to an afterglow in which the electron temperature rapidly cools to $T_e\sim0.25~\mathrm{eV}$ while the density decays exponentially with a time constant on the order of $10~\mathrm{ms}$. The LAPD is a pulsed device and the discharge is repeated with a $1~\mathrm{Hz}$ repetition rate for many consecutive hours; the plasma conditions are highly reproducible shot-to-shot and enable ensemble averaging during analysis. 

Located $15~\mathrm{m}$ from the BaO cathode are two $3$-mm-diameter crystal cathodes of cerium hexaboride (CeB$_6$) mounted on probe shafts inserted into the plasma; the crystals are supported by current-carrying wires mounted on a ceramic base that is affixed to the end of the probe shaft. Insulated wires inside the probe shafts connect the wire supports to outputs at the opposite end of the probe shafts located outside the LAPD vacuum chamber. CeB$_6$ has a low work function and high electron emissivity when it is heated to an operating temperature of around $1400~^\circ\mathrm{C}$; the heating is accomplished ohmically with around $10~\mathrm{W}$ of power to each crystal. To align the crystals in the LAPD at nearly the same location axially, one probe is inserted from the top and one from the side --- shown schematically in Fig.~\ref{fig:Schematic}; the axial location of the crystals is designated as $z_0=0~\mathrm{cm}$. The crystal inserted from the top is positioned near the centre of the machine and fixed for the duration of the experiments, while the crystal inserted from the side can be moved in the $(x,y)$ plane using a programmable probe drive with millimetre accuracy. 

A bias is applied between each crystal and the mesh anode starting $2~\mathrm{ms}$ into the afterglow phase; this time is designated as $t=0$, and the bias lasts for $20~\mathrm{ms}$ (see Fig.~\ref{fig:Schematic}. When the bias is applied, electrons are emitted from the CeB$_6$ crystals that rapidly thermalize in the afterglow plasma after a few mean free paths, creating a region of elevated temperature approximately $1~\mathrm{m}$ in extent and a few mm in diameter. The potential difference between each crystal and the anode can be varied, but is kept below $20~\mathrm{V}$ to ensure the energy of each emitted electron is below the ionization energy of helium ($24.6~\mathrm{eV}$). Magnetized plasmas have large anisotropy in parallel and transverse thermal transport coefficients; thus, the heated region in front of each crystal cathode rapidly forms a filamentary structure of elevated pressure several metres long with a symmetric Gaussian-like transverse profile less than $2~\mathrm{cm}$ in diameter. The peak temperature near the heat sources is $3$ to $5~\mathrm{eV}$, depending on the bias voltage, and decreases continuously toward the ends and edges of the filaments where the temperature equilibrates with the cold background plasma. 

The LAPD has access ports along the length of the vacuum chamber spaced $32$\;cm apart axially. The measurements presented herein are collected using small Langmuir probes inserted through the ports and biased to sample the temporal evolution of the current drawn from the plasma. Since the LAPD has a high repetition rate and high reproducibility, a probe located at an axial position $z$ can be placed at a position in the transverse $(x,y)$ plane, collect several nearly identical shots, and then be moved to a new position in the plane to repeat the process; this allows the collection of large 2D data planes from an ensemble of plasma shots. The process of moving the probes and collecting the data is entirely automated with the probes mounted on probe drives capable of less than $1$\;mm accuracy. In these experiments, probes are inserted at distances $z_1 = 256~\mathrm{cm}$ and $z_2 = 544~\mathrm{cm}$ from the crystal cathodes, as seen in Fig.~\ref{fig:Schematic}.

The Langmuir probes are used to collect two types of measurements: ion saturation current and temporal Langmuir sweeps. Ion saturation current, $I_\mathrm{sat}$, is collected at the probe face when the probe is biased well below the plasma potential; the measured current is proportional to the plasma density and square root of the electron temperature~\cite{Merlino2007}, i.e. $I_\mathrm{sat} \propto n\sqrt{T_e}$. The $I_\mathrm{sat}$ measurements are often decomposed into fluctuating and time-averaged components using digital filtering to remove frequency content below $1~\mathrm{kHz}$ from the $I_\mathrm{sat}$ signal to get $\delta I_\mathrm{sat}$. Temporal Langmuir sweeps can be analyzed to determine the electron temperature, $T_e$, plasma density, $n_e$, and space potential, $V_s$, from a characteristic $I$--$V$ curve~\cite{Chen2001}. The probe voltage is swept across a voltage range in a continuous sawtooth pattern with a period on the order of $200$ to $400$\;$\mu$s; each ``sweep" of the voltage can then be post-processed to extract the plasma parameters. This method is useful for acquiring information about the time-averaged plasma parameters but has poor temporal resolution and cannot yield information about fast fluctuations in the plasma. 

To highlight the edges of the filament where unstable modes can develop, gradient maps are taken using a central difference method,

\begin{align}
\| \nabla f(x_i,y_j) \| = \sqrt{f_x(x_i,y_j)^2+f_y(x_i,y_j)^2}\\
f_x(x_i,y_j)=\frac{f(x_{i+1},y_j) - f(x_{i-1},y_j)}{2 \delta}\\
f_y(x_i,y_j) = \frac{f(x_i,y_{j+1}) - f(x_i,y_{j-1})}{2 \delta}
\end{align}
where $i$ and $j$ denote the grid points and $\delta$ equals $1$\;mm. This is the spacing of the experimental measurement grid for all cases except the two filament close separation, which was taken on a $2$\;mm grid. In this particular case the data was linearly interpolated onto a $1$\;mm grid for consistency and to reduce the noise in the gradient maps.
%
\section{Results}\label{sec:ExpResults}

\subsection{Filament Evolution}

\begin{figure}[h]
{\includegraphics[trim= 16cm 2cm 14cm 0cm,clip,width=\linewidth]{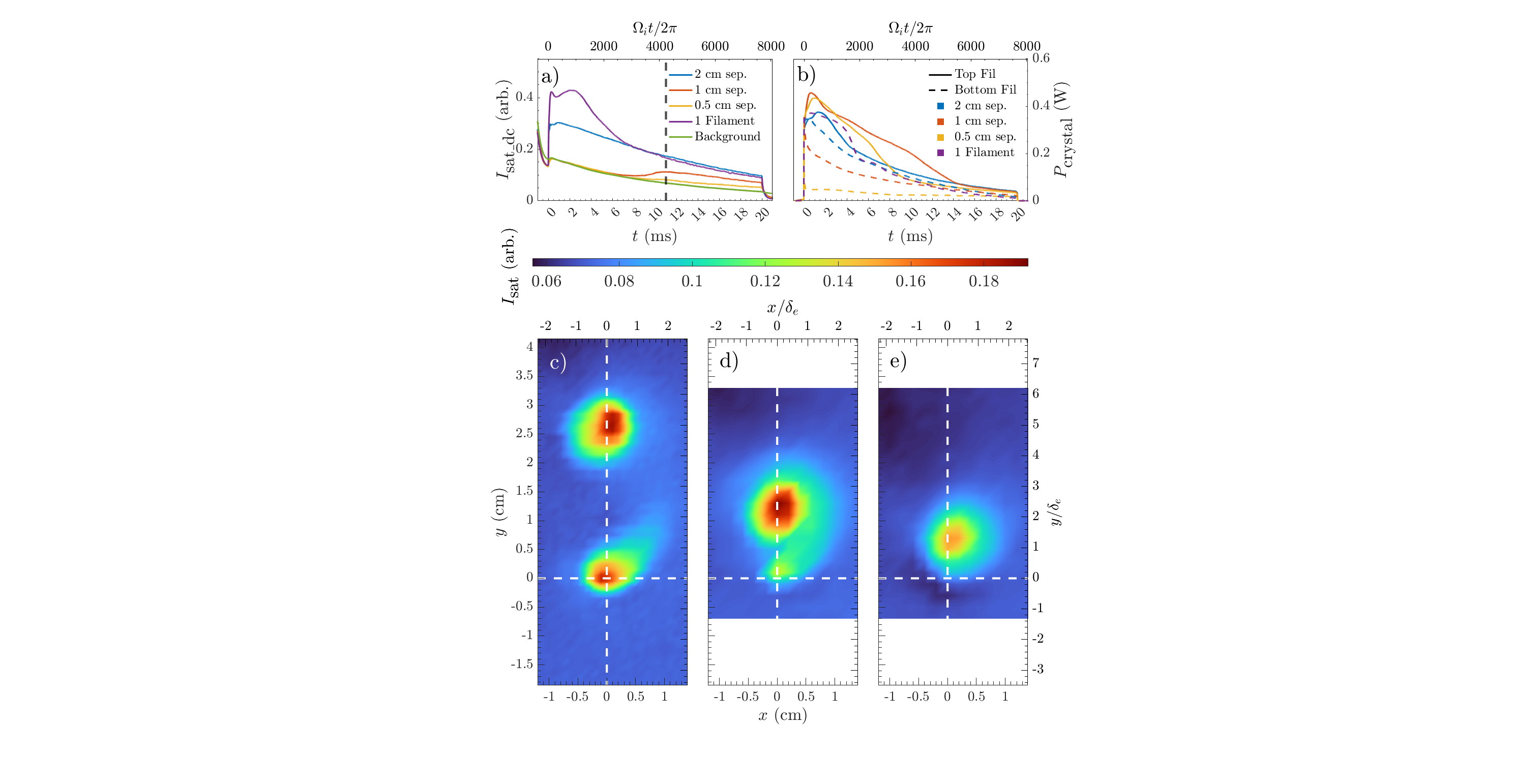}}
\caption{Panel a) shows the Evolution of $I_\textrm{sat}$ over time of the filament center for different conditions, and panel b) shows the top (solid line) and bottom (dashed line) filament power for difference spacings. Panels c) - e) show spatial planes of $I_\textrm{sat}$ taken at the dashed time in panel a) for the 2~cm, 1~cm, and 0.5~cm separations.  \label{fig:IsatTimePlane}}
\end{figure}

\begin{figure}[h!]
{\includegraphics[trim= 18cm 1cm 9cm 0cm,clip,width=\linewidth]{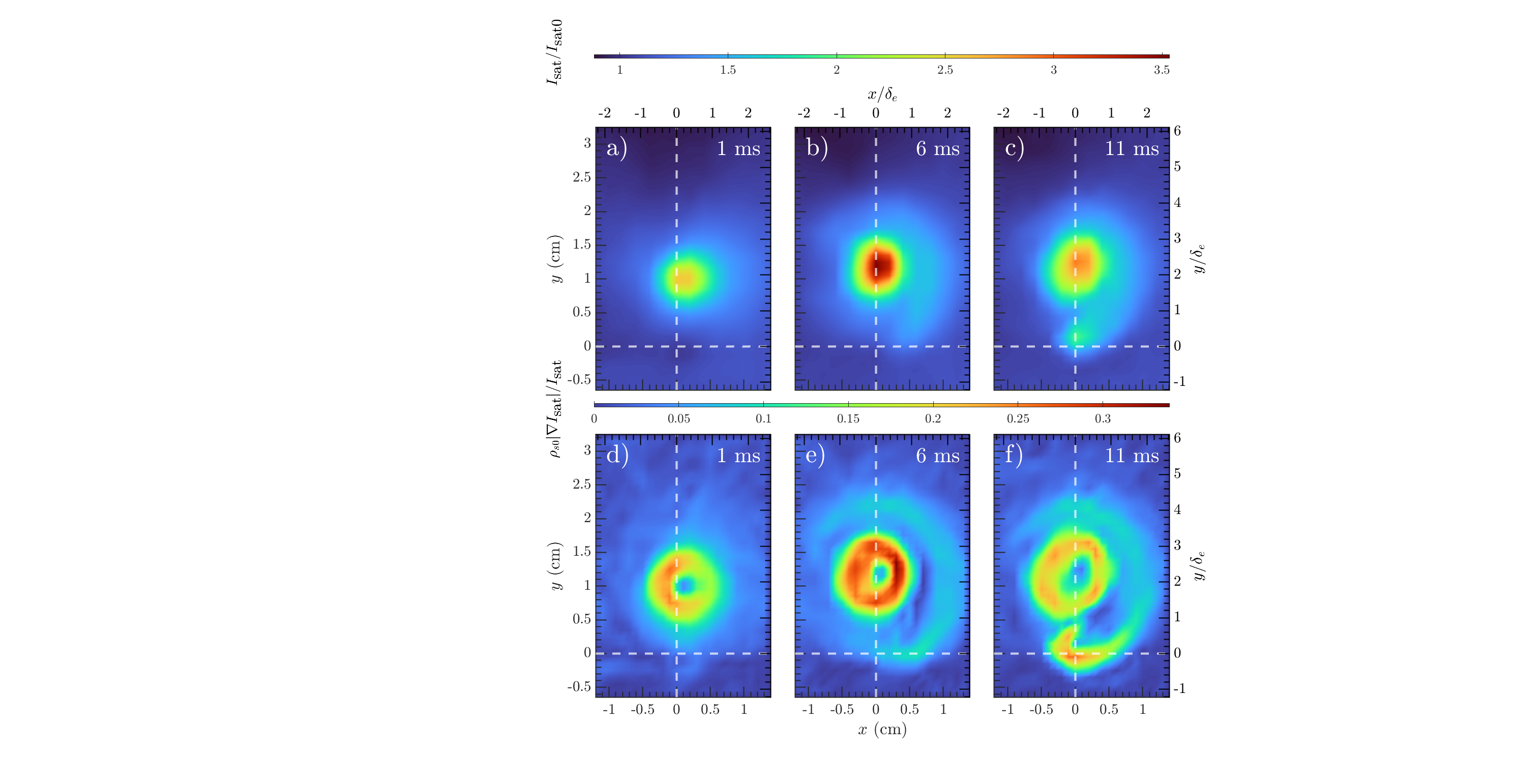}}
\caption{$I_\textrm{sat}$ and $\nabla I_\textrm{sat}$ for the 1~cm separation at 1~ms, 6~ms, and 11~ms. $I_\textrm{sat0}$ is the background $I_\textrm{sat}$ at the given time, taken as the average of the outer edge of the figure. The gradient maps highlight the wrapped tail region and the reformation of the bottom filament. \label{fig:GradIsat1cm}}
\end{figure}

Figure \ref{fig:IsatTimePlane} shows an overview of the plasma evolution. Panel a) shows the DC evolution of $I_\textrm{sat}$ at the coordinates x~=~0, y~=~0, which is set to be the location of the bottom filament. $t = 0$ corresponds to when the filaments are biased and begin emitting. The different plotted $I_\textrm{sat}$ traces show the measured $I_\textrm{sat}$ for different separations of the two emitting crystals. Additionally, traces are shown for only a single filament present, and with no filaments active to show the background plasma evolution. The single filament case shows significantly larger peak $I_\textrm{sat}$ values than the other conditions. In the 2~cm case, this is due to the formation of a tail region which draws temperature and density away from the central core of the filament, resulting in an approximately 25\% lower peak value. In the 1~cm case, the $I_\textrm{sat}$ curve follows the background until around 9~ms, where it starts to grow. This is due to the filaments combining on the top filament in the early parts of the emission with the bottom filament wrapped around the top filament. The bottom filament eventually reforms, but at reduced $I_\textrm{sat}$. A more detailed picture of the evolution for 1~cm separation throughout time is shown in Figure \ref{fig:GradIsat1cm}. In the 0.5~cm separation case, the filaments merge into a single larger filament positioned roughly on the location of the upper filament, and do not separate. Thus, there is no filament at the measured position, and the $I_\textrm{sat}$ is only slightly elevated from the background. Panel b) shows the filament power for the top (solid line) and bottom filament (dashed line) for the same cases. For the closer separations, emission from the bottom filament is suppressed due to shadowing, as the top filament crystal is physically located in front of the bottom filament within the machine. Panels c), d), and e) show a plane of $I_\textrm{sat}$ at the time indicated with the dashed line on panel a), which is 11~ms. Panel c) shows the formation of a tail on the bottom filament as discussed above. Panel d) shows a much stronger top filament and weaker bottom filament with an extended tail. In contrast, Panel e) shows only the upper filament as the filaments merge on the upper filament. A dual time axis is shown on Panels a) and b) in terms of the number of ion cyclotron revolutions, $\Omega_i t/2\pi$, and a secondary distance axis on panels c), d) and e) is shown in terms of the number of electron skin depths, $\delta_e$, calculated using a density of $10^{12} \textrm{cm}^{-3}$.

Spatial planes of electron temperature, density, and floating potential were extracted for the 2~cm separation by taking I-V curve sweeps over 0.5~ms. These sweeps can relatively noisy, and thus the plasma parameters are calculated by taking the average of 5 different shots. This allows the electron temperature and space potential to be extracted every 0.5~ms - via correlation the $I_\textrm{sat}$ data can be used along with the temperature to calculate the density.

\begin{figure}[h!]
{\includegraphics[trim= 20cm 1cm 18cm 0cm,clip,width=\linewidth]{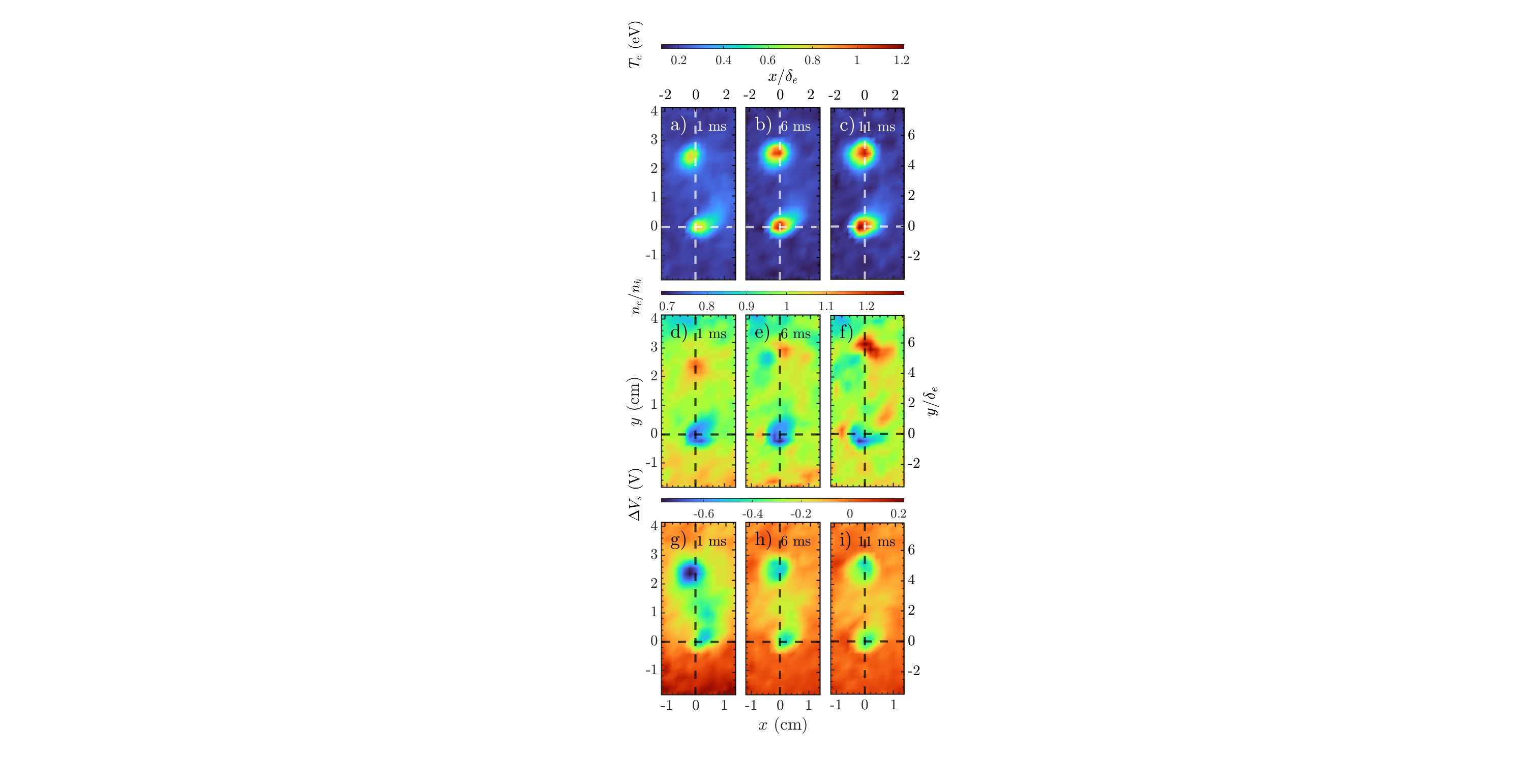}}
\caption{Maps of temperature, density, and potential for the 2~cm  at 1~ms, 6~ms, and 11~ms. Panels a)-c) show the filaments with a peak of roughly 1~eV compared with the background of roughly 0.2~eV. Panels d)-f) show the density normalized to background, and show the presence of a density depletion on the bottom filament and a density enhancement on the upper edge of the top filament in later timesteps. Panels g)-i) show the difference in space potential from the plane edge, which appears similar to the temperature distribution. \label{fig:TnVPanels}}
\end{figure}

In Figure \ref{fig:TnVPanels}, panels of the electron temperature $T_e$ (eV), density enhancement $n_e/n_b$, which is the electron density at the given position divided by the background density at that time, and the change in space potential with respect to the background space potential $\Delta V_s$ (V), are shown at three different times; 1~ms, 6~ms, and 11~ms. Panels a) - c) show the growth of the electron temperature as the filaments establish. The position of the bottom filament remains fixed, but a tail forms extending outwards towards the top filament. Panels d) - f), show the presence of a density depletion on the bottom filament and a slight density enhancement on the top filament, with the density enhancement significantly moving upwards and to the right in panels e) and f). Comparing the position of the density enhancement on panel f) with the location of the top filament in figure c), it is clear that they do not align. Also worth of note is the presence of  density enhancement on the outer tail region on the bottom filament, suggesting a channel of increased density between the filaments. This suggests density is being transported through flows from the tail region from the lower to the upper filament. Panels g), - i) show  potential wells associated with the filaments, and the upper well shifts to the right in the later panels. This closely matches the position of the filaments as seen in the temperature panels. Additionally, the depths of these wells decreases as time progresses. It is worth noting that the variation in electron temperature from approximately 0.2~eV in the background to 1.2~eV in the center of the filaments is a much wider variation than the variation in the background normalized density of 0.8 to 1.2. Thus, we describe the filaments as temperature filaments and the predominant variation in $I_\textrm{sat}$ is from the gradients in electron temperature, rather than variations in density. The gradients in temperature are particularly important as they lead to the growth of drift-Alfvén modes.

\begin{figure}[h]
{\includegraphics[trim= 4cm 2cm 1cm 0cm,clip,width=\linewidth]{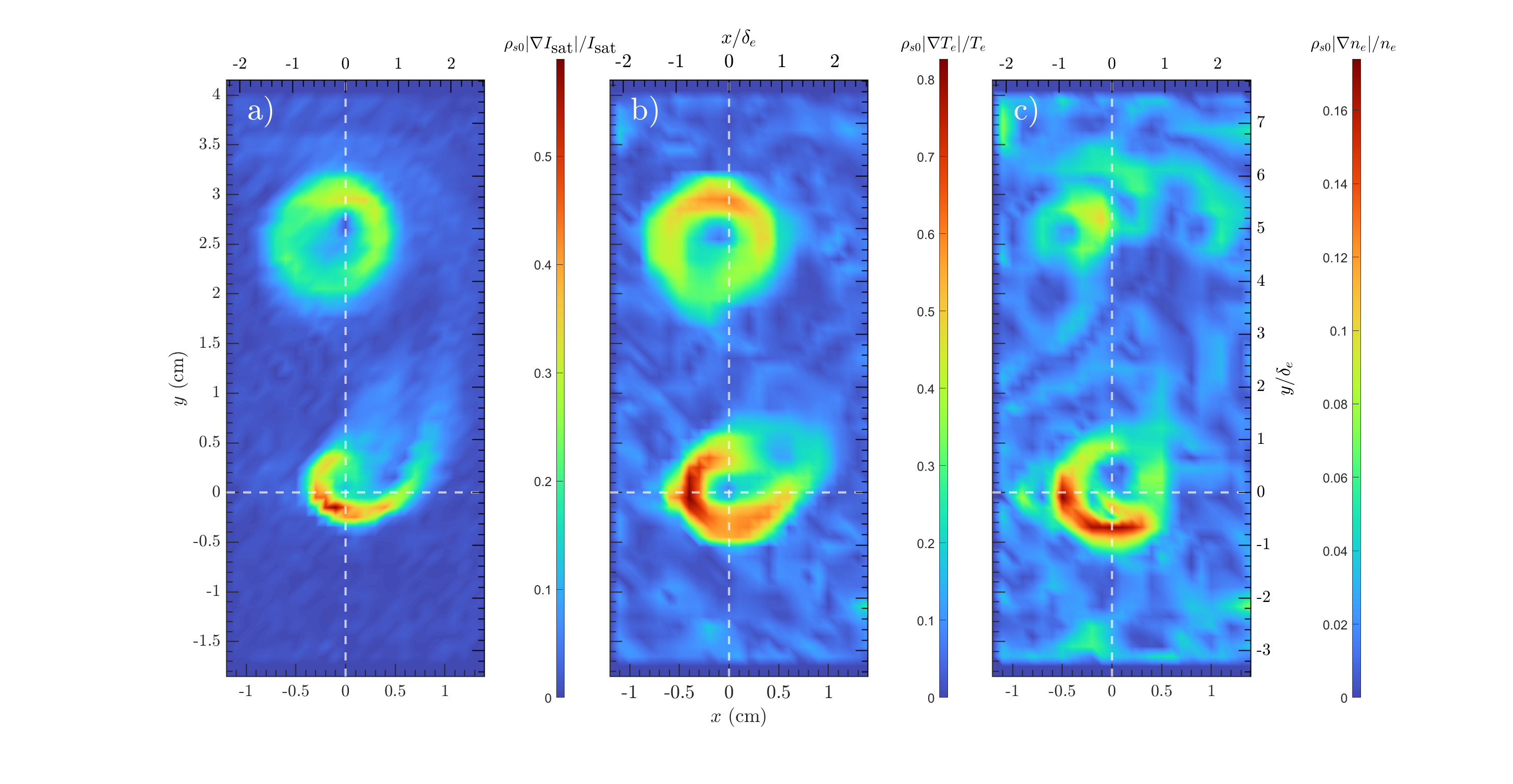}}
\caption{$\nabla I_\textrm{sat}$, $\nabla T_e$, and $\nabla n_e$ for the 2~cm separation, taken at $t = 6$. These maps highlight the gradient region and edges of the filaments, which are responsible for mode formation. \label{fig:GradIsatTn}}
\end{figure}

Figure \ref{fig:GradIsatTn} shows the normalized gradients at 6~ms in $I_\textrm{sat}$ (panel a) ), $T_e$ (panel b) ), and $n_e$ (panel c) ), multiplied by the ion sound length $\rho_{s0}$. This time corresponds to the center set of panels in Figure \ref{fig:TnVPanels}. Visualizing the gradients in these quantities highlights the tail region on the lower filament. There is a slight gradient in the $I_\textrm{sat}$ plot following the direction of the tail towards the upper filament on the outer edge as well. It is important to note that the direction of the gradient on the lower filament in panel c)  is opposite those in the other panels, as the center of the filament is a density depletion rather than a density enhancement. The temperature and $I_\textrm{sat}$ plots look much more similar than the density and $I_\textrm{sat}$ plots; this is because the relative variation in density between the maximum and minimum is fairly small compered with the temperature, which has a much larger variation. Thus, the contribution to variations in $I_\textrm{sat}$ from the temperature will largely outweigh that of the density. The x and y scales are shown both as absolute lengths and in terms of the number of electron skin depths.

For the 1~cm separation, $I_\textrm{sat}$ and gradient plots are presented in Figure \ref{fig:GradIsat1cm} at 3 different times to show the time-evolution of the filaments in the closer separation. The top panels are normalized to $I_\textrm{sat0}$ defined as the background $I_\textrm{sat}$ at each time step, and the bottom panels are the normalized gradient of $I_\textrm{sat}$ multiplied by the ion sound length. In panels a) and d), the filaments are shown at 1~ms, where there is only one visible filament in the position of the top filament and the bottom filament is absent, which is more evident in panel d). In panels b) and e), the filaments are shown at 6~ms. More clearly visible on the gradient plot, though still visible on panel b), is the presence of a ring of enhanced $I_\textrm{sat}$ wrapped around the central filament which corresponds to the second filament. In the gradient plot, the double-peaked radial structure is evident with a clear minima between them. At this time, no clear filament peak is evident on the bottom filament. In panels c) and f), the bottom filament is clearly visible in both plots, though weaker and smaller than the top filament. From the gradient plot, it is clear that the tail of this filament remains wrapped around the upper filament.

\begin{figure}[h]
{\includegraphics[trim= 16cm 1cm 16cm 0cm,clip,width=\linewidth]{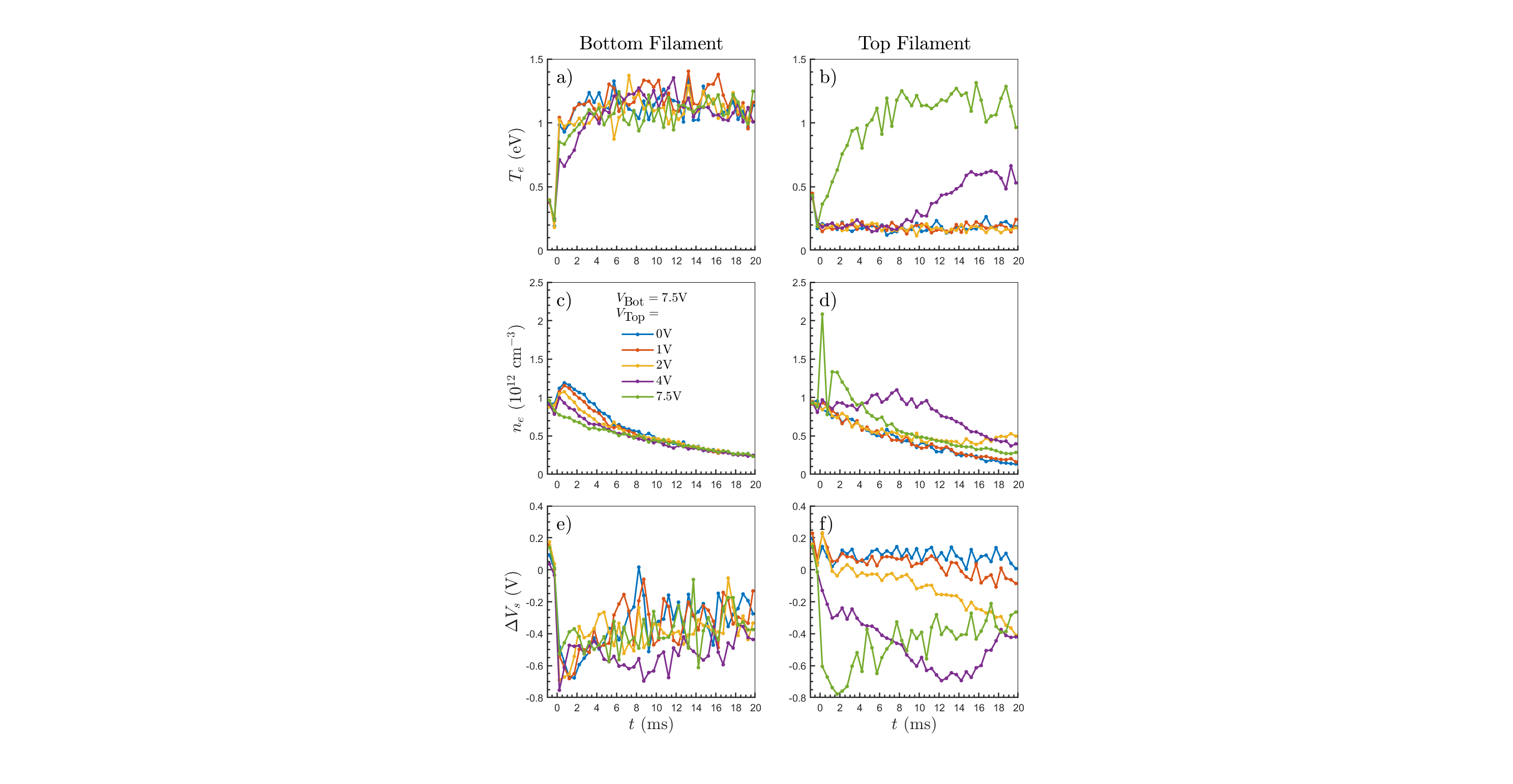}}
\caption{Time evolution of $T_e$, $n_e$, and $\Delta V_s$ for the 2~cm spacing case with different applied top filament voltages. Panels a), c) and e) are taken from the bottom filament, while panels b), d) and f) are taken from the top filament. \label{fig:TimeEvTnV}}
\end{figure}

Data was also taken with reduced applied voltages on the top filament, with the bottom filament always remaining at 7.5~V. In Figure \ref{fig:TimeEvTnV}, time traces of $T_e$ on panels a) and b), $n_e$ on panels c) and d), and $\Delta V_s$, defined as the difference between the space potential and background potential on panels e) and f) are shown for varying bottom filament voltage. Panels a), c), and e) are taken at x~=~0~mm y=~-0.5 ~mm, centered on the bottom filament, and panels b), d), and f) are taken at x~=~0~mm y~=~25.5~mm, centered on the top filament. Panel a) shows the bottom filament temperatures, which remain roughly the same (within noise), suggesting that the top filament voltage does not significantly impact the temperature of the bottom filament. In panel b), the temperature varies greatly with the applied voltage, with no clear filament established in any cases but the 7.5V and 4V case, and with the 4V case taking longer to establish. In panel c), the effect of tail formation and of density being pulled into the tail is evident, as the higher top filament voltage (corresponding to a larger top filament and hence greater tail formation), the lower the density seen in the early stages of filament evolution. The slow decay in all cases is due to the decaying background density. In panel d), we see that the higher voltages for which filaments are established, particularly in the  7.5V, and 4V cases, with the spike in density occurring earlier for the 7.5V case than the 4V case. This likely could corresponds to the filament establishing itself earlier at higher filament power (connected with the temperature plots). There appears to be a slight rise in density in the 2V case, but this occurs very late in the time series. It is worth noting that the increase in density in each case occurs before the rise in temperature. Panels e) and f) show the fluctuation in space potential relative to background. For the top filament in panel f), the 7.5V case roughly matches the traces in panel e) for the bottom filament, while the dip in the 4V case seems to match an inverted combination of the temperature and density enhancements.

\begin{figure}[h]
{\includegraphics[trim= 12cm 4cm 12cm 4cm,clip,width=\linewidth]{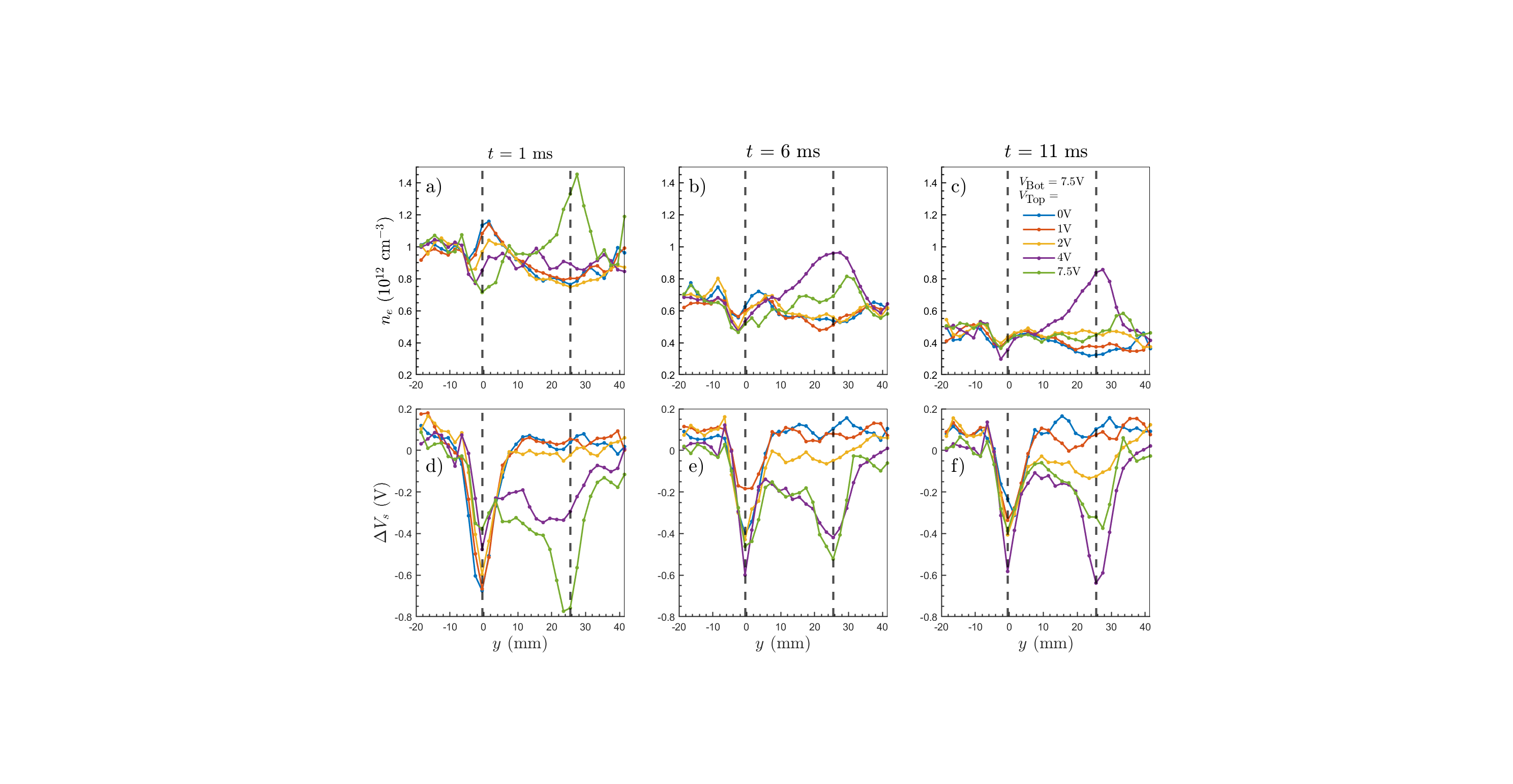}}
\caption{Line cuts along the y axis for $n_e$ and $\Delta V_s$, taken at $x=0$ for differing top filament voltages at 1~ms, 6~ms, and 11~ms. The dashed lines indicate the locations of the top and bottom filaments. Panels a)-c) show $n_e$ - the dropping background level with increased time is due to the background plasma density dropping in the afterglow. Panels d)-f) show the difference in space potential from background - the deep well that forms in the 4~V case is particularly notable. \label{fig:ySlicesnV}}
\end{figure}

Finally, Figure \ref{fig:ySlicesnV} shows line cuts along the y axis taken through the center of the filaments for density and space potential at t = 1~ms, 6~ms, and 11~ms. Temperature profiles were not shown, as these were very similar across different time steps with a sharp peak similar to the plane in Figure \ref{fig:TnVPanels}. The dashed lines indicate the locations of the filaments (also corresponding to the location taken in Figure \ref{fig:TimeEvTnV}). The filament shown to the left is the bottom filament. In panel a), the bottom filament shows slight density enhancements on the bottom (left) filament for the low voltage cases, and density depletions on the higher voltage cases. There is a clear and fairly broad peak of enhanced density on the top filament for the 7.5V case, while the other filaments remain roughly at or slightly below background. By t = 6~ms (in panel b) ), there is a large density enhancement on the top filament for the 4V case, a smaller density enhancement on the 7.5V case, and a slight depletion in the other cases. There also seems to be a density depletion below the bottom filament for all voltages.
In panel c) the density enhancement is even more drastic on the 4V case, and appears to have shifted on the 7.5V case, matching what was seen in Figure \ref{fig:TnVPanels}. Panels d), e), and f) show the potentials. Panel d) mirrors what is seen in panel a) with the density, with the density depletion corresponding to a shallower potential well on the bottom filament. In panel e), a large dip in the potential of the top filament in the 4V case is seen, which grows even larger in panel f) with the larger spike in density in c), showing good correlation between the potential and density.

\subsection{Mode Structure and Coherence}

\begin{figure}[h]
{\includegraphics[trim= 13cm 3cm 6cm 0cm,clip,width=\linewidth]{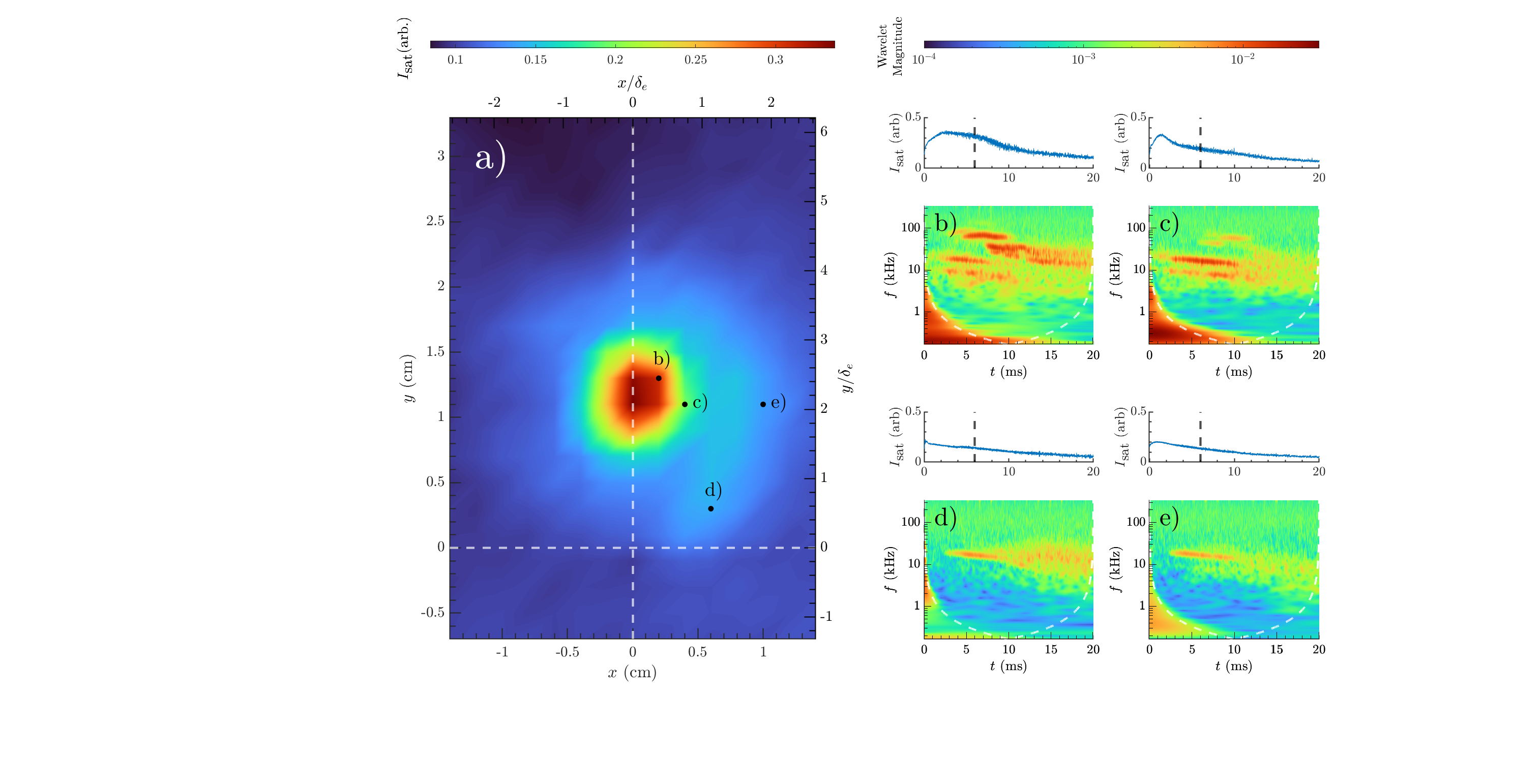}}
\caption{Plane of $I_\textrm{sat}$ at t = 6~ms with reference locations as black dots. Panels b)-e) show a single shot of $I_\textrm{sat}$ at the chosen location as well as the wavelet transform at each of the reference locations (averaged over shots). The dashed line in the time traces at 6~ms corresponds to the time chosen for panel a). The wavelet transforms show the dominant mode frequencies for each of the reference locations. \label{fig:wavelets1cm}}
\end{figure}

Next, an examination of the mode structure and coherence will be presented to examine the degree of coupling between the filaments and filament interaction, starting with the 1-filament case. Figure \ref{fig:wavelets1cm} shows a panel of $I_\textrm{sat}$ taken at 6~ms as panel a), as well as points indicating reference locations b) - e) which correspond to panels b-e) on the right. Location b) is inside the “main” filament on the upper-right edge, position c) is located on the right edge of the main filament gradient, position d) is located in the tail region of the bottom filament which will re—emerge in later time steps, and position e) is located on the exterior edge of the outside gradient/tail gradient. These positions were selected to show a representative frequency spectrum of the overall frequency activity in both the upper and re-emerging lower filament. Panels b)-e) consist of a single-shot time trace at the reference location, and the (shot-averaged) wavelet transform of the $I_\textrm{sat}$ signal. The area below the dashed white line includes edge-effects and should be ignored. Panels b) and c) show that power at many frequencies are present simultaneously, with main modes in the early time steps at around 10kHz and 2f components in the filament center (panel b) and edge. The higher frequency components at ~80kHz are mainly localized to the filament center and are difficult to analyze due to insufficient spatial resolution. Panels d) and e) show that the exterior gradient, which is shallower, only supports the 2f component at ~20kHz. The slow drop in mode frequency over time is expected and a result of background density decay throughout the afterglow.

\begin{figure}[h]
{\includegraphics[trim= 15cm 3cm 14cm 0cm,clip,width=\linewidth]{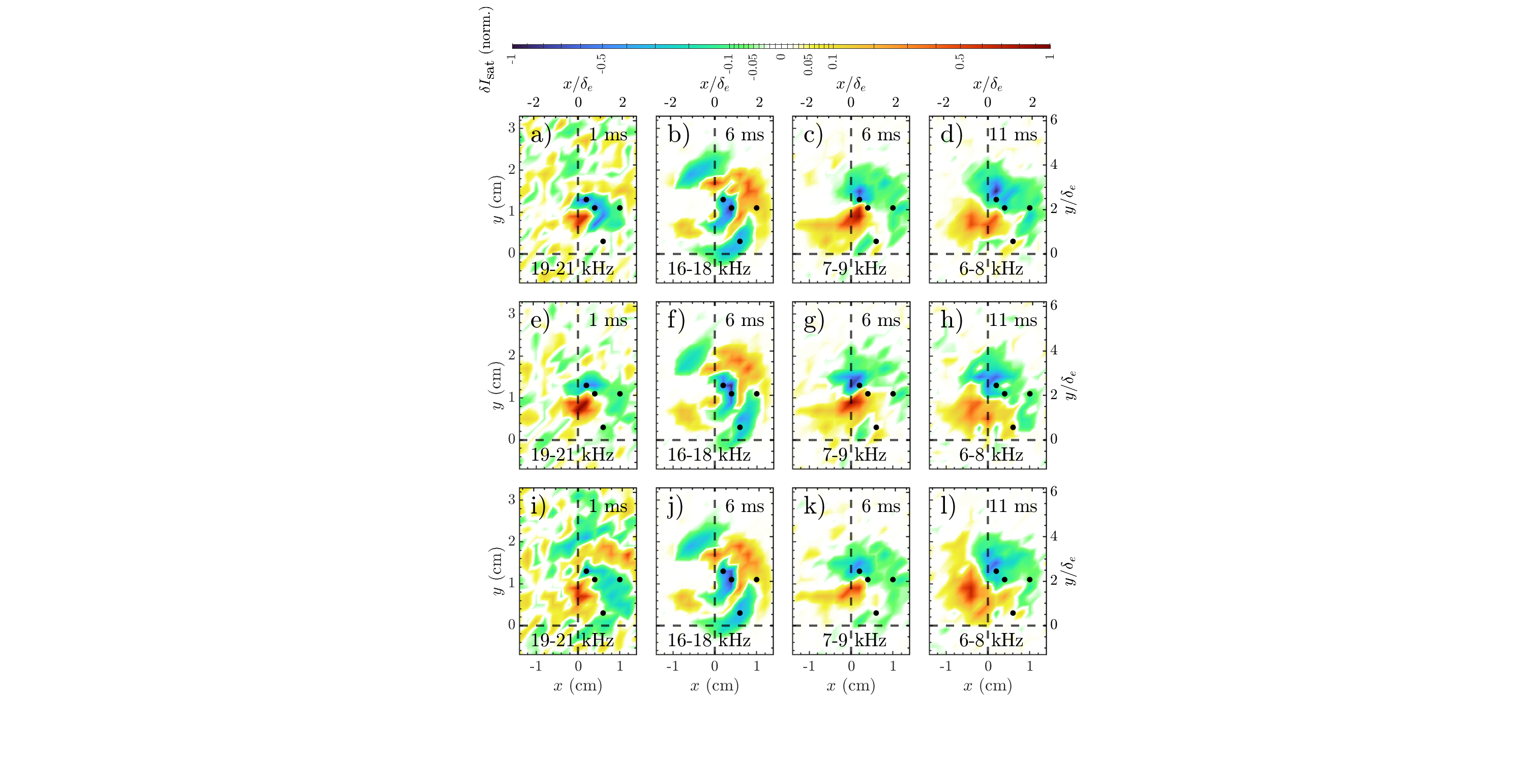}}
\caption{Reconstruction of $\delta I_\textrm{sat}$ via cross power/cross phase analysis using the top crystal current for panels a)-d), reference probe for e-h, and bottom crystal current for i-l). The modes are shown on a 2-ended log scale which is linear around zero. \label{fig:modes1cm}}
\end{figure}

To examine the spatial extent of the modes, Fourier transforms were taken over 1~ms windows starting at the standard 1~ms, 6~ms, 11~ms spacing used in this paper. These mods are shown in Figure \ref{fig:modes1cm} filtered around frequencies of interest. However, as the modes are not coherent shot-to-shot, a reference phase for each shot must be used, which is done by multiplying by the conjugate of the Fourier transform of the reference signal for each shot. In this set of experiments, 3 separate sources of reference signal are used. The standard method of using a reference probe positioned 96~cm from the measurement plane (352~cm from the emitter) is shown in panels e-h). This is compared with using the crystal currents as reference signals - panels a-d) for the top crystal (top row of panels) and panels i-l) for the bottom crystal. Here, each of the rows are generally quite similar, because the system is very coupled with wrapped filaments. The similarities in mode structure regardless of reference source primarily demonstrates the validity of using any of the reference methods. In the far separation case, the reference methods will no longer produce the same results for the top and bottom crystal currents (and no frequency content was available on the reference probe in this experiment, likely indicating a misplaced reference probe) - this will be discussed further in the next section. 

The mode structures shown here for one instantaneous value of $\delta I_\textrm{sat}$ are functionally equivalent to taking the real component (or any other complex angle) of the Fourier transform. This $\delta I_\textrm{sat}$ is averaged over a $2$~kHz window to improve signal-to-noise. It is important to note that each individual $ \delta I_\textrm{sat}$ panels shown here is advanced in time/rotated in complex phase angle to best highlight the mode structure and to show a consistent mode picture between the difference reference sources. The locations of the reference points from the wavelet transforms in the previous figure are shown as solid dots to provide positional reference. Panels a-b) show the 2f mode, which displays a m~=~2 mode structure strongest near the inner filament but present on both the interior and exterior gradients. The lower frequency mode is shown in panels c-d) and displays a m=1 structure which also extends through the interior and exterior gradients.

\begin{figure}[h]
{\includegraphics[trim= 14cm 3cm 6cm 0cm,clip,width=\linewidth]{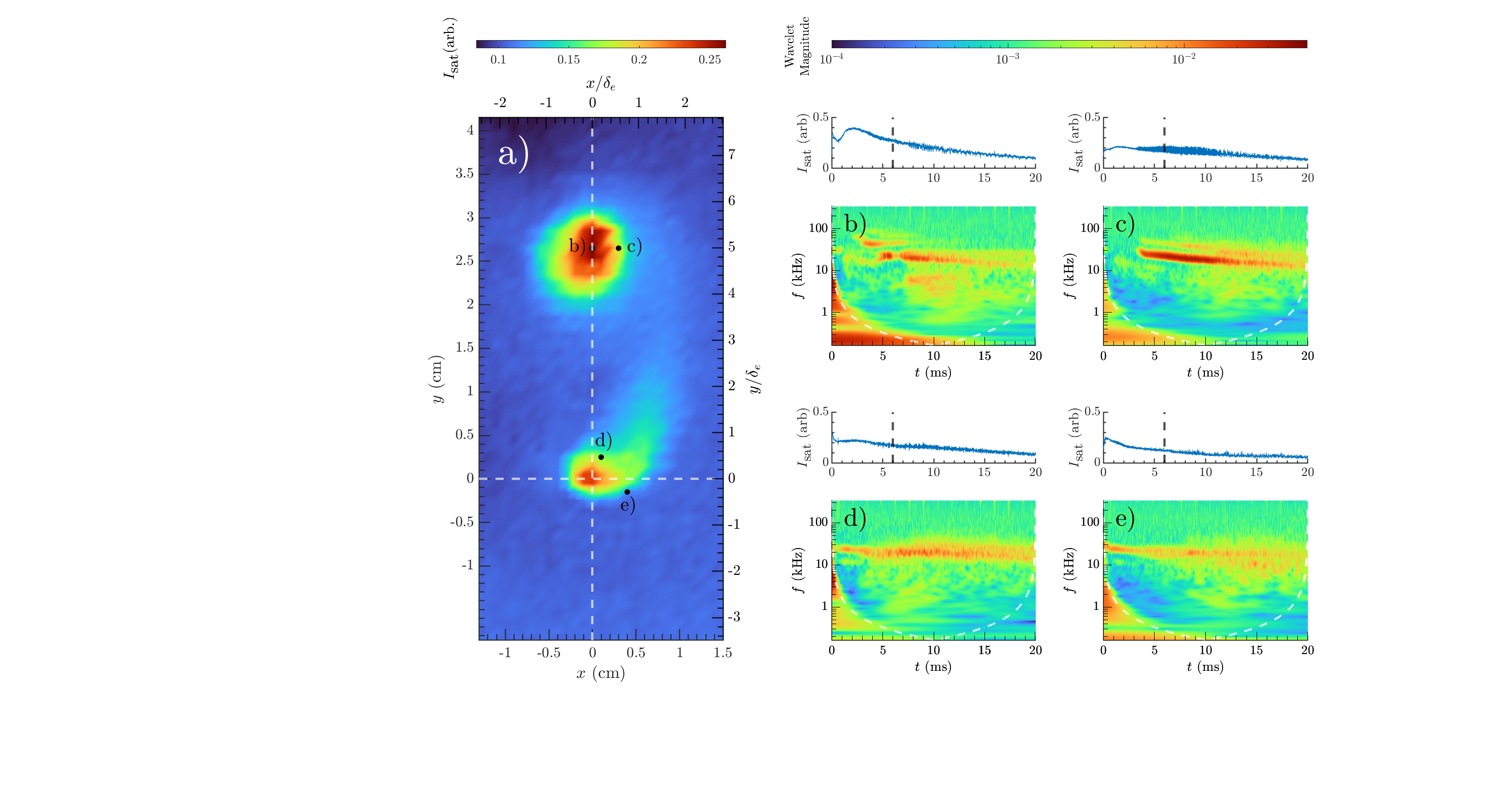}}
\caption{Plane of $I_\textrm{sat}$ at t = 6~ms with reference locations as black dots. Panels b)-e) show a single shot of $I_\textrm{sat}$ at the chosen location as well as the wavelet transform at each of the reference locations (averaged over shots). The dashed line in the time traces at 6~ms corresponds to the time chosen for panel a). The wavelet transforms show the dominant mode frequencies for each of the reference locations. \label{fig:wavelets2cm}}
\end{figure}

Next, the 2~cm separation case will be discussed. Figure \ref{fig:wavelets2cm} shows a panel of $I_\textrm{sat}$ taken at 6~ms as panel a), as well as points indicating reference locations b) - e) which correspond to panels b-e) on the right. Location b) is on the interior of the upper filament, Location c) is on the edge of the upper filament, location d is on the interior edge of the lower filament tail, and location e) is on the outer edge of the filament tail. Location b) shows some higher order modes in early time steps, but eventually settles into the $\sim$20~kHz mode. Location c) shows the dominant $\sim$20~kHz mode starting at about 4~ms, as well as a double frequency component at 40~kHz. The lower filament locations show consistent mode activity in the 20~kHz range throughout the time series. It is interesting that the $\sim$10~kHz mode present in the 1~cm separation case does not appear to develop in the 2~cm case.

\begin{figure}[h!]
{\includegraphics[trim= 16cm 1cm 14cm 0cm,clip,width=\linewidth]{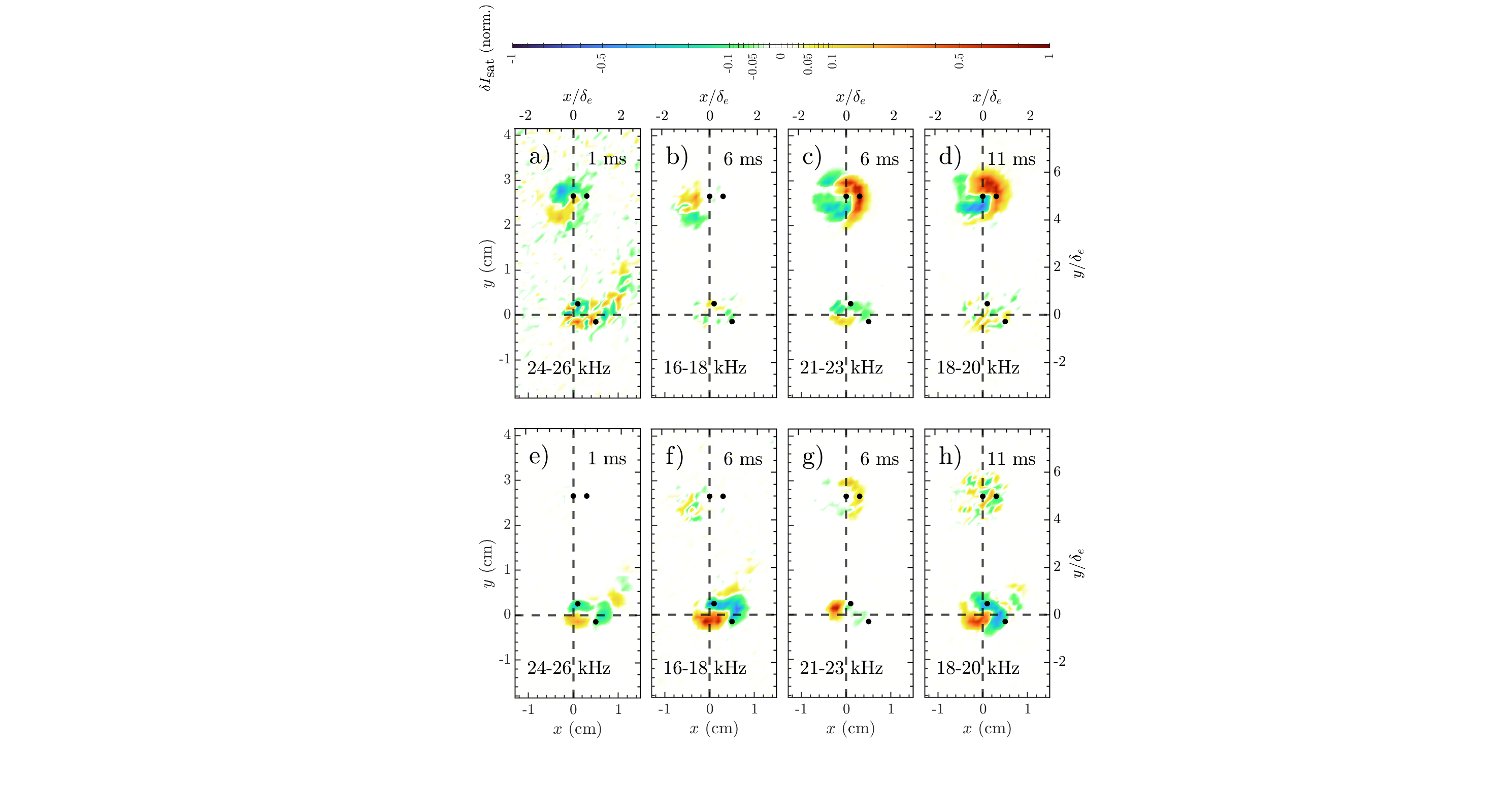}}
\caption{Reconstruction of $\delta I_\textrm{sat}$ via cross power/cross phase analysis using the top crystal current for panels a)-d) and bottom crystal current for e-h). The modes are shown on a 2-ended log scale which is linear around zero. \label{fig:modes2cm}}
\end{figure}

By taking the Fourier transform and correlating with a reference source, the mode structure for the 2~cm case can be reconstructed. This is shown in Figure \ref{fig:modes2cm}. Unfortunately, this dataset did not have a reference probe with frequency information, so the only available reference sources are the top and bottom currents. As compared with Figure \ref{fig:modes1cm}, this is equivalent to omitting the middle row. Panels a-d) use the top crystal reference current, and panels e-h) use the bottom crystal reference current. The dots indicate the reference locations from Figure \ref{fig:wavelets2cm}. Panel e), taken at 1~ms with the bottom filament as reference current, shows a clear mode structure on the bottom filament extending out into the filament tail. There is no apparent mode structure on the top filament. Panel a), taking the top current as reference, shows a weak mode structure on the top filament and frequency activity on the bottom filament but does not show a clear mode structure on the bottom filament like panel e). Panels b)/f) and c)/g are taken at 6~ms over different frequency ranges - 16-18~kHz is the main frequency band of the bottom filament mode while 21-23~kHz is the main frequency band of the top filament. Panel f) clearly shows the bottom filament mode, while panel b) does not and only shows frequency activity on the left side of the top filament. As the probes are inserted from the right side of the plane, measurements in this location require the probe to be inserted through the bulk of the filament and thus it is probable these measurements are not a real effect. In comparison, panel c) shows a clear mode on the top filament and some weak frequency activity on the bottom filament, while panel g) shows only weak mode structure in both locations. This second set of panels at 6~ms are selected from the frequency range for the top filament, so these results are not unexpected. By 11~ms, shown in panels d) and h), both the top and bottom filaments are in similar enough frequency ranges that the top mode is visible when correlating with the top current, and the bottom mode is visible when correlating with the bottom current. It is particularly interesting to note the extension of the mode onto the tail of the bottom filament.

\begin{figure}[h!]
{\includegraphics[trim= 0cm 0cm 0cm 0cm,clip,width=\linewidth]{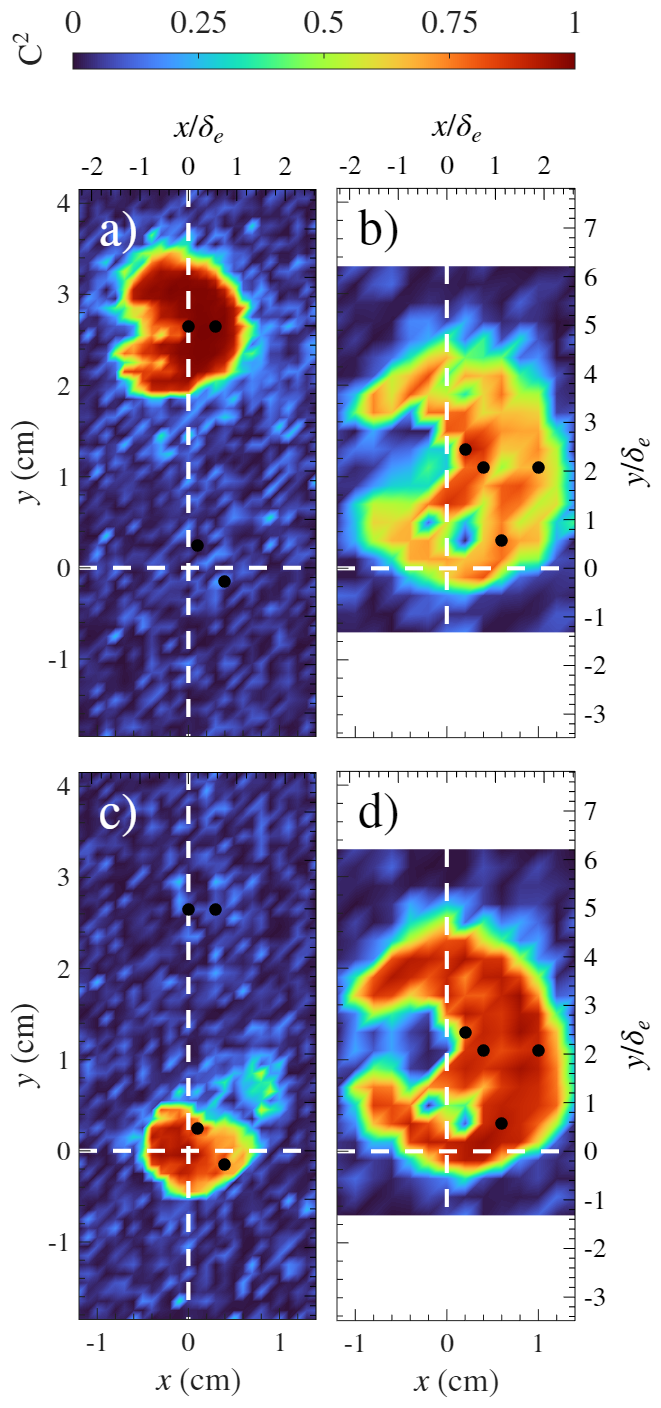}}
\caption{Coherence between $I_\textrm{sat}$ and reference current for 1~cm and 2~cm spacing at 8~ms over the 18~kHz to 20~kHz range. Panels a) and b) are the coherence with the top filament current while panels c) and d) are the coherence with the bottom filament current. The dots are the reference locations from Figures \ref{fig:wavelets1cm} and \ref{fig:wavelets2cm}.  \label{fig:1cm2cmCoherPlanes}}
\end{figure}

To quantify the degree of coupling between filaments, the coherence between $I_\textrm{sat}$ and the filament current will be used. The coherence is given by Equation \ref{eq:Coher}

\begin{align}
C^2 = \frac{\left| \langle F^{ }_{I\_\textrm{sat}} F_{I\_\textrm{ref}}^*  \rangle\right|^2}   {\langle \left|F_{I\_\textrm{sat}}\right| \left|F_{I\_\textrm{ref}}  \right|\rangle^2}
    \label{eq:Coher}
\end{align}

where $C^2$ is the coherence, which ranges from 0 to 1, $F^{ }_{I\_\textrm{sat}}$ denotes the complex valued Fourier transform of $I_\textrm{sat}$, $F_{I\_\textrm{ref}}$ denotes the complex valued Fourier transform of the reference current (either top or bottom), $^*$ denotes complex conjugation, and the angle brackets $\langle~~~~\rangle$ denote averaging. To improve signal to noise ratio, this averaging is conducted both over shots and by sub-windowing the Fourier transform (8 to 8.25~ms, 8.25 to 8.5~ms, 8.5 to 8.75~ms, and 8.75 to 9~ms). 8~ms was chosen for the 2~cm case because the mode frequencies of the top and bottom filaments overlap at this time, giving the best chance for coherence between the top and bottom filaments. The denominator takes the absolute value before shot-averaging, while the numerator takes the shot average of the complex-valued product, then takes the absolute value. In this way, if the shot-to-shot phase of the product is not consistent, the numerator will approach 0, and in the limit of perfect phase coupling the coherence will approach 1. Figure \ref{fig:1cm2cmCoherPlanes} shows 2~cm and 1~cm coherence as a function of distance. Both separations are taken over the frequency range 18~kHz to 20~kHz - it is a coincidence that the dominant mode frequency is the same for each.  The black dots represent the locations where wavelet transforms were taken in Figures \ref{fig:wavelets1cm} and \ref{fig:wavelets2cm}. Panels a) and b) are taken using the top current, while panels c) and d) are taken using the bottom current. Panels a) and c) show that in the 2~cm case, the filament modes are not coupled with each-other and each filament is only coherent with its respective current. In comparison, panels b) and d) show that the 1~cm case is a fully coupled system and the modes are coherent across the whole wrapped-filament structure. Interestingly, panel b) shows more coherence in the center-core where the top filament is present compared to panel d), but panel d) shows more coherence along the outer edge of the wrapped filament-tail.

\begin{figure}[h]
{\includegraphics[trim= 4cm 6cm 32cm 1.8cm,clip,width=\linewidth]{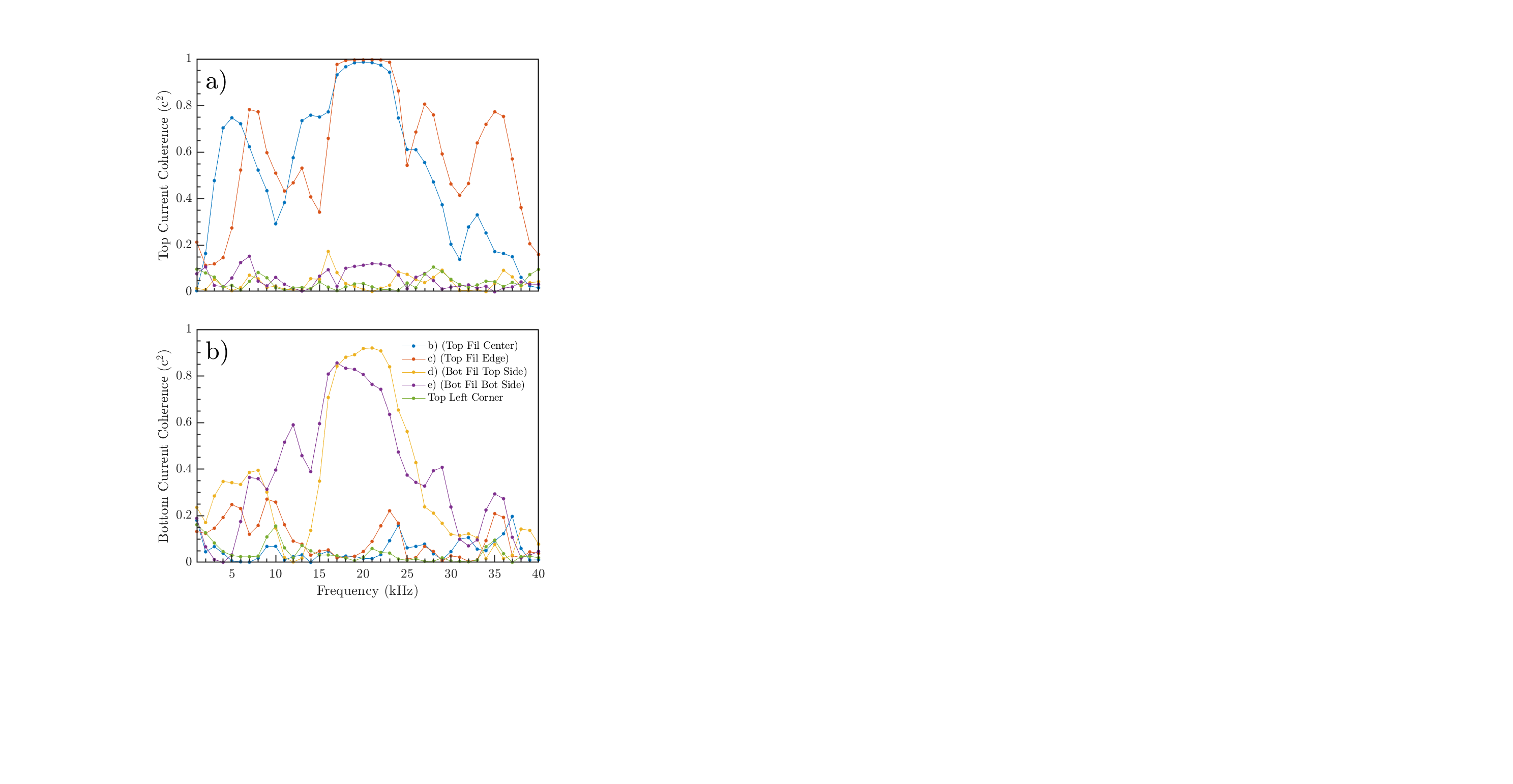}}
\caption{Coherence between $I_\textrm{sat}$ and reference current at 8~ms (7-9ms in total) as a function of frequency. The locations b)-e) referred to are the reference locations from Figure \ref{fig:wavelets2cm}. \label{fig:1cm2cmCoherTraces}}
\end{figure}

Figure \ref{fig:1cm2cmCoherTraces} shows coherence as a function of frequency for the 2~cm spacing. Panel a) shows the coherence for various locations with the top crystal current, while panel b) shows the coherence with the bottom crystal current. The letters associated with each reference location in the legend are the reference point locations from Figures  \ref{fig:wavelets2cm}. Additionally, the top left corner of the data plane is included as a reference for the coherence seen in the "background" far away from the filaments. The 2~cm top current coherence reference shows the "top filament" associated references - locations b) and c) - as very coherence in the $\sim$16-26~kHz range. In contrast, the bottom current coherence shows strong coherence in the mode frequency band with both lower filament locations and minimal coherence elsewhere. There appears to be no coupling between filaments regardless of the frequency range taken, further demonstrating that the 2-filament case is uncoupled between the top and bottom filaments.

%
\section{Discussion}\label{sec:Discussion}
This section contains a discussion of the filament merging dynamics and the transport of vorticity. Since this a low beta plasma the current density is too weak for merging via parallel current attraction.

\begin{figure}[h]
{\includegraphics[trim= 18cm 1cm 11cm 0cm,clip,width=\linewidth]{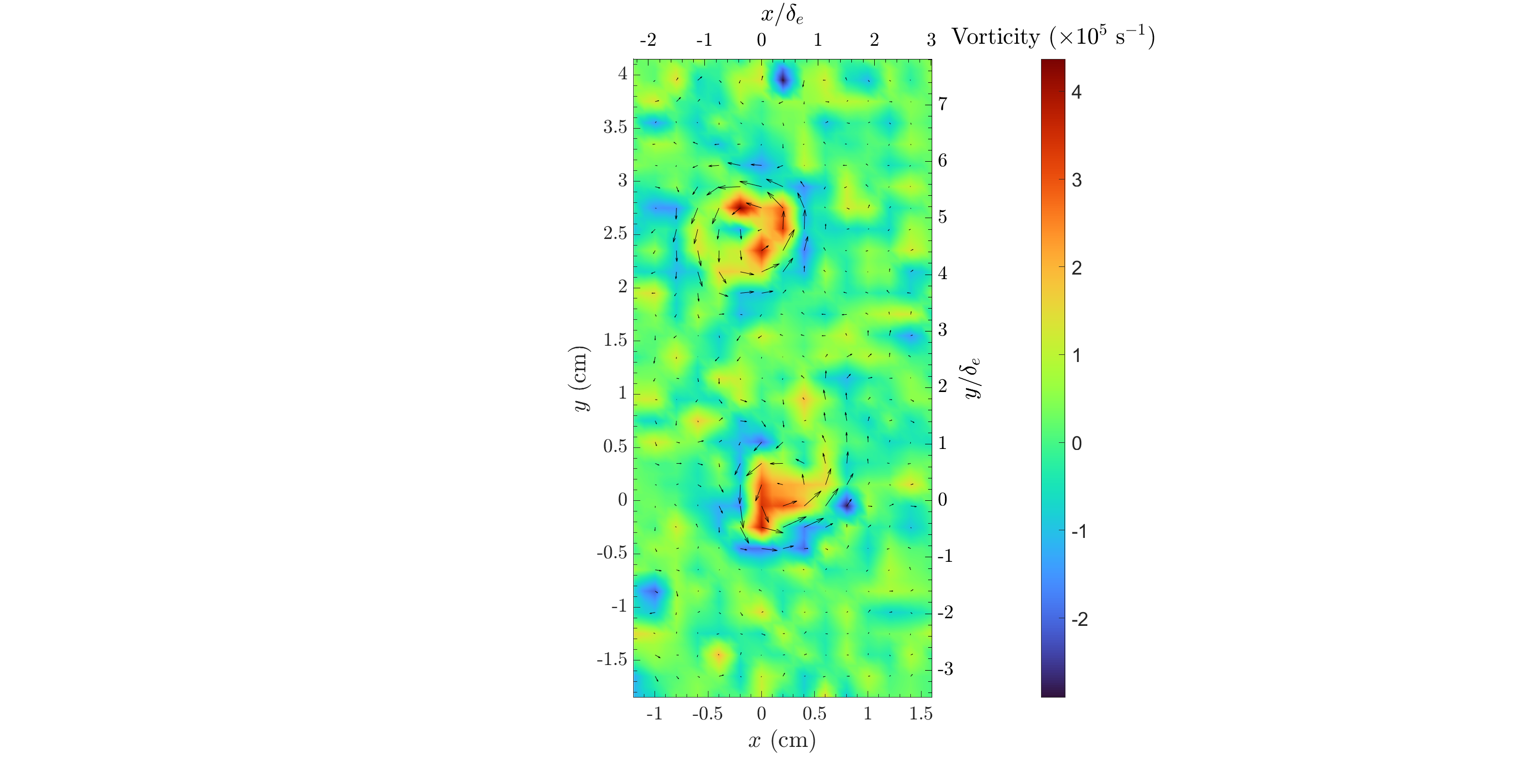}}
\caption{Vorticity and $E \times B$ flow for the 2~cm spacing. The colormap shows the vorticity calculated using equation \ref{eq:Vort}, while the arrows show the direction and relative magnitude of the $E \times B$ flow with B as the background magnetic field and $E$ calculated from $V_s$. \label{fig:cExBandVort}}
\end{figure}

In the individual magnetized temperature filaments there is a circular $\bf{E} \times \bf{B}$ plasma flow (m=0) along with electric field fluctuations from the drift-Alfv\'{e}n modes (finite m). The vorticity is advected by this flow. As the two like-sign vortices become closer there is a differential rotation and stretching of the vorticity distribution, creating a vortex tail. Figure \ref{fig:cExBandVort} shows the vorticity as the colormap and $E \times B$ flows as the arrows for the 2~cm separation case. The arrows are scaled according to the relative magnitude of the $E \times B$ flow at each location and are located with the tail of the arrow at the gridpoint they correspond to. $B$ is a experimental parameter of the setup - 1000 gauss, oriented out of the plane, and $E$ is calculated from $-\nabla V_s$ from I-V sweeps. $V_s$ is shown in Figure \ref{fig:TnVPanels}. The vorticity is defined in equation \ref{eq:Vort}, where the last equality is only true for $V_s$ in the x-y plane and $B$ along the z axis.
\begin{align}
\bf{\omega} = \nabla \times \bf{v} = \nabla \times \left(\frac{E\times B}{B^2}\right) = \frac{\nabla^2 V_s}{B}
    \label{eq:Vort}
\end{align}

The vorticity and arrows are taken at 6~ms, averaged over a 0.5~ms window, which corresponds to two I-V sweeps (one at 5.75~ms, one at 6.25~ms). The direction of the $\bf{E} \times \bf{B}$ flow shows the transport of material through the tail of the bottom filament. Both filament cores have a positive vorticity, with a small ring of negative vorticity surrounding them.

The antisymmetric part of the vorticity field associated with the overlapping outer filaments drives convective mass and energy transfer between the vortex-filament cores pulling them together. The presence of drift-Alfv\'{e}n modes smooths out the sharp vorticity gradients at the merging boundary, and when sufficiently close, there is a final coalescence of the core vorticity.

%
%
\section{Summary}\label{sec:Summary}

The main findings from the experiments on the interaction 
of two magnetized electron temperature filaments in a low beta plasma
regime are summarized as follows: \\

(i) Close separation

\begin{itemize}
\item When close, asymmetric filament power leads to one filament wrapping around the other due to a deeper potential well, but this weaker filament later re-emerges.

\item From the global mode structure, the amplitude is peaked on the internal, stronger filament and the external (wrapped tail) gradients with fully coherent fluctuations across the global mode.
\end{itemize}

(ii) Far separation

\begin{itemize}
\item Further apart, asymmetry leads to the formation of a convective tail which transports density from the weaker to the stronger filament, thus establishing a weak polarization electric field between the filaments.

\item Modes remain decoupled at further separation with each filament having its own separate drift-Alfv\'{e}n modes which are not phase-locked.
\end{itemize}

Finally, from an analysis of the vorticity distribution and $\bf{E} \times \bf{B}$ flow, significant vorticity transport occurs along with stretching, accompanied by drift-Alfv\'{e}n fluctuations, each playing a role in the final merging dynamics.

%
%
%
\begin{acknowledgments}
The authors acknowledge support from the Natural Sciences and Engineering Research Council of Canada (NSERC). The experiments were performed at the Basic Plasma Science Facility supported by DOE and NSF, with major facility instrumentation developed via an NSF award AGS-9724366.
\end{acknowledgments}
%
%
\bibliography{References.bib}

@PREAMBLE{
 "\providecommand{\noopsort}[1]{}" 
 # "\providecommand{\singleletter}[1]{#1}%" 
}

@article{Sydora2019,
	Author      = {Sydora, R. D. and Karbashewski, S. and Van Compernolle, B. and Poulos, M. J. and Loughran, J.},
    Title       = {Drift-Alfv{\'e}n fluctuations and transport in multiple interacting magnetized electron temperature filaments},
	Journal     = {J. Plasma Phys.},
	Volume      = {85},
    Pages       = {905850612},
	Year        = {2019},
    Url         = {https://doi.org/10.1017/S0022377819000886}
	}

@article{Burke1998,
	Author      = {Burke, A. T. and Maggs, J. E. and Morales, G. J.},
	Title       = {Observation of Simultaneous Axial and Transverse Classical Heat Transport in a Magnetized Plasma},
	Journal     = {Phys. Rev. Lett.},
    Volume      = {81},
	Pages       = {3659},
	Year        = {1998},
	Url         = {https://doi.org/10.1103/PhysRevLett.81.3659}
	}

@article{DIppolito2011,
author = {D'Ippolito, D. A. and Myra, J. R. and Zweben, S. J.},
doi = {10.1063/1.3594609},
issn = {1070664X},
journal = {Physics of Plasmas},
number = {6},
title = {{Convective transport by intermittent blob-filaments: Comparison of theory and experiment}},
volume = {18},
year = {2011}
}

@article{Zweben2002,
author = {Zweben, S. J. and Stotler, D. P. and Terry, J. L. and Labombard, B. and Greenwald, M. and Muterspaugh, M. and Pitcher, C. S. and Hallatschek, K. and Maqueda, R. J. and Rogers, B. and Lowrance, J. L. and Mastrocola, V. J. and Renda, G. F.},
doi = {10.1063/1.1445179},
issn = {1070664X},
journal = {Physics of Plasmas},
number = {5},
pages = {1981--1989},
title = {{Edge turbulence imaging in the Alcator C-Mod tokamak}},
volume = {9},
year = {2002}
}

@article{Boedo2003,
author = {Boedo, J. A. and Rudakov, D. L. and Moyer, R. A. and McKee, G. R. and Colchin, R. J. and Schaffer, M. J. and Stangeby, P. G. and West, W. P. and Allen, S. L. and Evans, T. E. and Fonck, R. J. and Hollmann, E. M. and Krasheninnikov, S. and Leonard, A. W. and Nevins, W. and Mahdavi, M. A. and Porter, G. D. and Tynan, G. R. and Whyte, D. G. and Xu, X.},
doi = {10.1063/1.1563259},
issn = {1070664X},
journal = {Physics of Plasmas},
number = {5 II},
pages = {1670--1677},
title = {{Transport by intermittency in the boundary of the DIII-D tokamak}},
volume = {10},
year = {2003}
}

@article{Carter2006,
author = {Carter, T. A.},
doi = {10.1063/1.2158929},
issn = {1070664X},
journal = {Physics of Plasmas},
number = {1},
pages = {1--4},
title = {{Intermittent turbulence and turbulent structures in a linear magnetized plasma}},
volume = {13},
year = {2006}
}

@book{Krasheninnikov2008,
author = {Krasheninnikov, S. I. and D'Ippolito, D. A. and Myra, J. R.},
booktitle = {Journal of Plasma Physics},
doi = {10.1017/S0022377807006940},
isbn = {0022377807006},
issn = {00223778},
number = {5},
pages = {679--717},
title = {{Recent theoretical progress in understanding coherent structures in edge and SOL turbulence}},
volume = {74},
year = {2008}
}

@article{Furno2008,
  title = {Experimental Observation of the Blob-Generation Mechanism from Interchange Waves in a Plasma},
  author = {Furno, I. and Labit, B. and Podest\`a, M. and Fasoli, A. and M\"uller, S. H. and Poli, F. M. and Ricci, P. and Theiler, C. and Brunner, S. and Diallo, A. and Graves, J.},
  journal = {Phys. Rev. Lett.},
  volume = {100},
  issue = {5},
  pages = {055004},
  numpages = {4},
  year = {2008},
  month = {Feb},
  publisher = {American Physical Society},
  doi = {10.1103/PhysRevLett.100.055004},
  url = {https://link.aps.org/doi/10.1103/PhysRevLett.100.055004}
}

@article{Bisai2019,
author = {Bisai, N. and Banerjee, Santanu and Sen, Abhijit},
doi = {10.1063/1.5082241},
issn = {10897674},
journal = {Physics of Plasmas},
number = {2},
title = {{A universal mechanism for plasma blob formation}},
url = {http://dx.doi.org/10.1063/1.5082241},
volume = {26},
year = {2019}
}

@article{Theiler2011,
author = {Theiler, C. and Furno, I. and Fasoli, A. and Ricci, P. and Labit, B. and Iraji, D.},
doi = {10.1063/1.3562944},
issn = {1070664X},
journal = {Physics of Plasmas},
number = {5},
title = {{Blob motion and control in simple magnetized plasmas}},
volume = {18},
year = {2011}
}

@article{Terry2017,
author = {Terry, J. L. and Ballinger, S. and Brunner, D. and LaBombard, B. and White, A. E. and Zweben, S. J.},
doi = {10.1016/j.nme.2016.11.020},
issn = {23521791},
journal = {Nuclear Materials and Energy},
pages = {989--993},
publisher = {Elsevier Ltd},
title = {{Fast imaging of filaments in the X-point region of Alcator C-Mod}},
url = {https://doi.org/10.1016/j.nme.2016.11.020},
volume = {12},
year = {2017}
}

@article{Pierre2004,
author = {Pierre, Th and Escarguel, A. and Guyomarc'h, D. and Barni, R. and Riccardi, C.},
doi = {10.1103/PhysRevLett.92.065004},
issn = {10797114},
journal = {Physical Review Letters},
number = {6},
pages = {13--16},
title = {{Radial Convection of Plasma Structures in a Turbulent Rotating Magnetized-Plasma Column}},
volume = {92},
year = {2004}
}

@article{Offeddu2022,
author = {Offeddu, N. and W{\"{u}}thrich, C. and Han, W. and Theiler, C. and Golfinopoulos, T. and Terry, J. L. and Marmar, E. and Galperti, C. and Andrebe, Y. and Duval, B. P. and Bertizzolo, R. and Clement, A. and F{\'{e}}vrier, O. and Elaian, H. and G{\"{o}}nczy, D. and Landis, J. D.},
doi = {10.1063/5.0126398},
issn = {10897623},
journal = {Review of Scientific Instruments},
number = {12},
pmid = {36586925},
publisher = {AIP Publishing, LLC},
title = {{Gas puff imaging on the TCV tokamak}},
url = {https://doi.org/10.1063/5.0126398},
volume = {93},
year = {2022}
}

@article{Angus2012,
author = {Angus, Justin R. and Umansky, Maxim V. and Krasheninnikov, Sergei I.},
doi = {10.1103/PhysRevLett.108.215002},
issn = {00319007},
journal = {Physical Review Letters},
number = {21},
pages = {1--5},
title = {{Effect of drift waves on plasma blob dynamics}},
volume = {108},
year = {2012}
}

@article{Easy2014,
author = {Easy, L. and Militello, F. and Omotani, J. and Dudson, B. and Havl{\'{i}}{\v{c}}kov{\'{a}}, E. and Tamain, P. and Naulin, V. and Nielsen, A. H.},
doi = {10.1063/1.4904207},
issn = {10897674},
journal = {Physics of Plasmas},
number = {12},
title = {{Three dimensional simulations of plasma filaments in the scrape off layer: A comparison with models of reduced dimensionality}},
url = {http://dx.doi.org/10.1063/1.4904207},
volume = {21},
year = {2014}
}

@article{Windisch2006,
author = {Windisch, T. and Grulke, O. and Klinger, T.},
doi = {10.1063/1.2400845},
issn = {1070664X},
journal = {Physics of Plasmas},
number = {12},
title = {{Radial propagation of structures in drift wave turbulence}},
volume = {13},
year = {2006}
}

@article{Antar2007,
author = {Antar, G. Y. and Yu, J. H. and Tynan, G.},
doi = {10.1063/1.2424886},
issn = {1070664X},
journal = {Physics of Plasmas},
number = {2},
title = {{The origin of convective structures in the scrape-off layer of linear magnetic fusion devices investigated by fast imaging}},
volume = {14},
year = {2007}
}

@article{Nielsen1996,
author = {Nielsen, A. H. and P{\'{e}}cseli, H. L. and {Juul Rasmussen}, J.},
doi = {10.1063/1.872008},
issn = {1070664X},
journal = {Physics of Plasmas},
number = {5},
pages = {1530--1544},
title = {{Turbulent transport in low-$\beta$ plasmas}},
volume = {3},
year = {1996}
}

@article{Karbashewski2022,
author = {Karbashewski, S. and Sydora, R. D. and {Van Compernolle}, B. and Simala-Grant, T. and Poulos, M. J.},
doi = {10.1063/5.0104283},
issn = {10897674},
journal = {Physics of Plasmas},
number = {11},
publisher = {AIP Publishing LLC},
title = {{Magnetized plasma pressure filaments: Analysis of chaotic and intermittent transport events driven by drift-Alfv{\'{e}}n modes}},
volume = {29},
year = {2022}
}

@article{Gekelman2016,
author = {Gekelman, W. and Pribyl, P. and Lucky, Z. and Drandell, M. and Leneman, D. and Maggs, J. and Vincena, S. and {Van Compernolle}, B. and Tripathi, S. K.P. and Morales, G. and Carter, T. A. and Wang, Y. and DeHaas, T.},
doi = {10.1063/1.4941079},
issn = {10897623},
journal = {Review of Scientific Instruments},
number = {2},
title = {{The upgraded Large Plasma Device, a machine for studying frontier basic plasma physics}},
url = {http://dx.doi.org/10.1063/1.4941079},
volume = {87},
year = {2016}
}

@article{Garcia2004,
author = {Garcia, O. E. and Naulin, V. and Nielsen, A. H. and Rasmussen, J. Juul},
doi = {10.1103/PhysRevLett.92.165003},
issn = {00319007},
journal = {Physical Review Letters},
number = {16},
pages = {4--7},
title = {{Computations of intermittent transport in scrape-off layer plasmas}},
volume = {92},
year = {2004}
}

@article{Zweben2016,
author = {Zweben, S. J. and Myra, J. R. and Davis, W. M. and D'Ippolito, D. A. and Gray, T. K. and Kaye, S. M. and Leblanc, B. P. and Maqueda, R. J. and Russell, D. A. and Stotler, D. P.},
doi = {10.1088/0741-3335/58/4/044007},
issn = {13616587},
journal = {Plasma Physics and Controlled Fusion},
number = {4},
publisher = {IOP Publishing},
title = {{Blob structure and motion in the edge and SOL of NSTX}},
volume = {58},
year = {2016}
}

@article{Chen2001,
	Author = {Chen,Francis F.},
	Journal = {Phys. Plasmas},
	Number = {6},
	Pages = {3029},
	Title = {Langmuir probe analysis for high density plasmas},
	Url = {https://doi.org/10.1063/1.1368874},
	Volume = {8},
	Year = {2001}}

@article{Merlino2007,
	Author = {Merlino, R. L.},
	Doi = {10.1119/1.2772282},
	Journal = {Am. J. Phys.},
	Number = {12},
	Pages = {1078},
	Title = {Understanding Langmuir Probe Current-Voltage Characteristics},
	Url = {https://doi.org/10.1119/1.2772282},
	Volume = {75},
	Year = {2007}}

@article{Sydora2024,
    author = {Sydora, R. D. and Simala-Grant, T. and Karbashewski, S. and Jimenez, F. and Van Compernolle, B. and Poulos, M. J.},
    title = {Experiments and gyrokinetic simulations of the nonlinear interaction between spinning magnetized plasma pressure filaments},
    journal = {Physics of Plasmas},
    volume = {31},
    number = {8},
    pages = {082304},
    year = {2024},
    month = {08},
    issn = {1070-664X},
    doi = {10.1063/5.0213345,}
}

%
%
\end{document}